\documentclass[10pt,letterpaper, twocolumn]{article}   
\usepackage{ol2}
\usepackage{graphicx}
\usepackage[draft]{hyperref} 
\begin{document}
\twocolumn[

\title{Precision spectral manipulation of optical pulses using a coherent photon echo memory}


\author{B.~C.~Buchler$^*$, M.~Hosseini, G.~H\'{e}tet, B.~M.~Sparkes and P.~K.~Lam}
\address{ARC Centre of Excellence for Quantum-Atom Optics,  Department of Quantum Science, \\
The Australian National University, Canberra ACT 0200, Australia}
\address{$^*$Corresponding author: ben.buchler@anu.edu.au}

\begin{abstract} 
Photon echo schemes are excellent candidates for high efficiency coherent optical memory. They are capable of high-bandwidth multi-pulse storage, pulse resequencing and have been shown theoretically to be compatible with quantum information applications.  One particular photon echo scheme is the gradient echo memory (GEM). In this system, an atomic frequency gradient is induced in the direction of light propagation leading to a Fourier decomposition of the optical spectrum along the length of the storage medium. This Fourier encoding allows precision spectral manipulation of the stored light.  In this letter, we show frequency shifting, spectral compression, spectral splitting, and fine dispersion control of optical pulses using GEM.  
\end{abstract}

]

Spectral manipulation is important for pulse compression, sideband extraction, and matching of pulse spectra to resonant and spectroscopic systems. Wideband manipulation, for ultrafast lasers and telecommunications applications, can be achieved via prisms and gratings.  Precision, narrowband manipulation can be achieved via optical high-Q optical resonances. In this letter, we propose an alternative narrowband manipulation technique where we imprint the pulse spectrum into an ensemble of atoms
that can be spectrally manipulated. This allows highly flexible frequency filtering and precision control of the output pulse spectrum.

Photon echo schemes have some very attractive properties for optical storage.  They have been shown to have GHz bandwidth \cite{Merkel} and are capable of storing many pulses \cite{LinOL1995}.  They can be used for pulse compression \cite{wangOL2000}, pulse shaping \cite{BarberOE02} and have been shown to preserve optical coherence. In particular Controlled Reversible Inhomogeneous Broadening (CRIB) systems \cite{Alex-PRL} have been shown to be compatible with quantum state storage.

The gradient echo memory (GEM) is a variant of CRIB that relies on the inhomogeneous broadening being introduced as an atomic frequency gradient along the length of the storage medium.  GEM has recently been demonstrated in both two \cite{HetetPRL} and three-level \cite{OL, nature09} atomic systems. Modelling has shown that it is ideally 100\% efficiency \cite{HetetPRL,Longdell08pra} and demonstrations using warm atomic vapours have so far reached 42\% efficiency\cite{nature09}.

\small
\begin{figure}
 \includegraphics[width=\columnwidth]{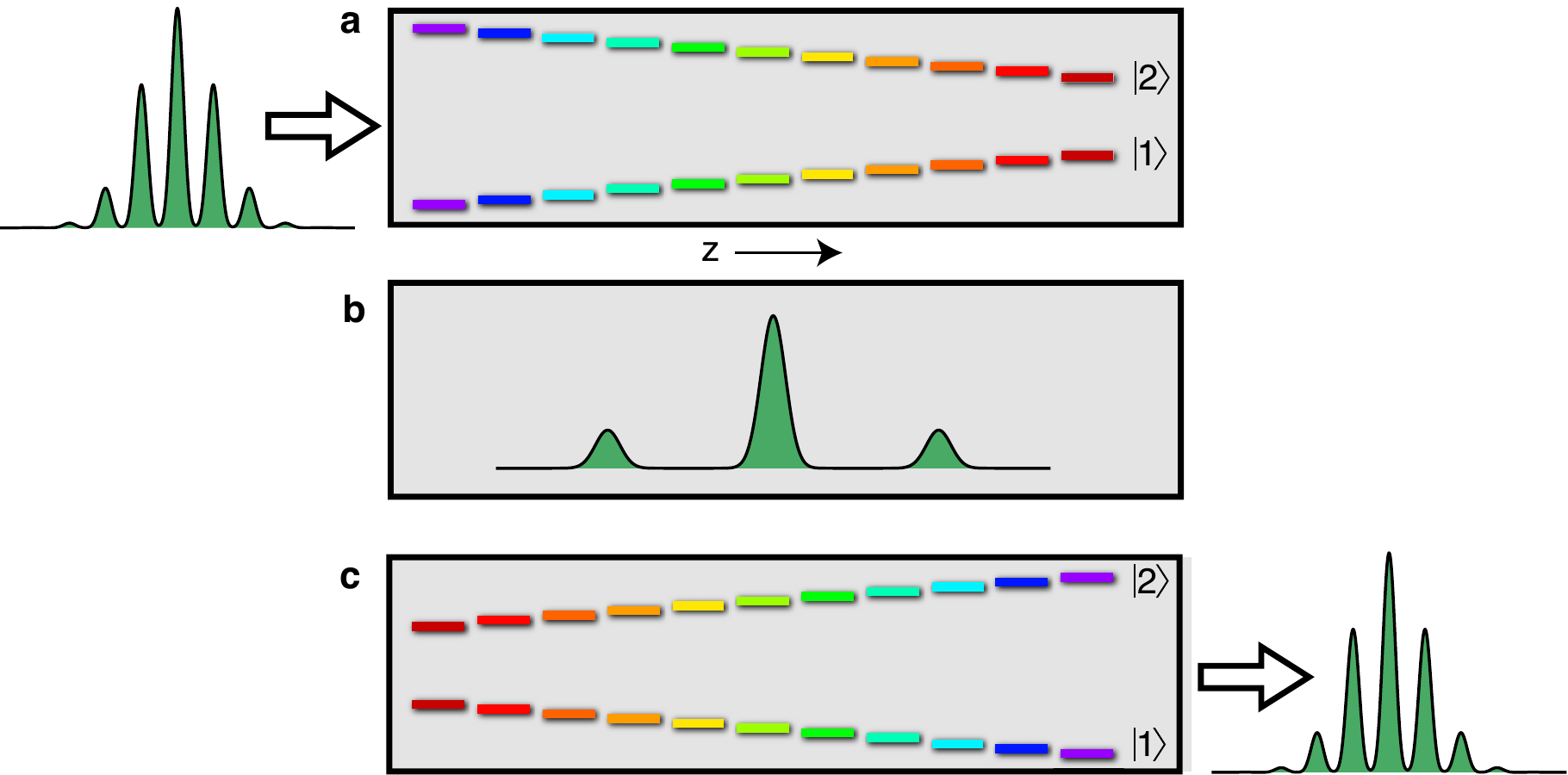}
 \caption{ a: The atomic storage medium with an atomic frequency spectrum ($\eta z$).  The bandwidth of the atomic broadening covers the input modulated pulse spectrum. b:  The Fourier spectrum of the input pulse is absorbed and stored as an atomic polarisation. c: To release the pulse, $\eta$ is switched to $-\eta$ at time $\tau$ and the stored light emerges at time $2 \tau$.}
  \label{schematic}
  \end{figure}

The GEM protocol works without any need for $\pi$-pulses. Rephasing is controlled by the linear atomic frequency spectrum, $\eta z$, that is induced along the length of the storage medium. Consider, for example, an ensemble of two-level atoms as shown in Fig.~\ref{schematic}(a).  Each frequency of the probe optical pulse ($E$) is absorbed by the ensemble at a different point along the length of the memory and is stored in the atomic polarisation ($\alpha$). The atomic polarisation along the $z$ direction is thus proportional to the Fourier spectrum of $E$ as shown in Fig.~\ref{schematic}(b). To release the stored light, the gradient $\eta$ is simply inverted at time $\tau$ and the optical field is regenerated as a photon echo at time $2\tau$, as shown in Fig.~\ref{schematic}(c). In this most basic two-level scheme the input pulses emerge in the forward direction but shape-reversed \cite{GEMpolariton}.  Using the extra degree of freedom afforded by a three-level system, the input pulses can be recalled without reversal or even in an arbitrary order \cite{nature09}.

We now consider manipulating the atomic frequency gradient to allow fine-control of the optical field recall. Rather than just inverting a simple linear gradient, we can invert different parts of the gradient at different times, add shifts to the atomic frequency spectrum and change the recall gradient by invoking a more general atomic frequency spectrum $\eta(t,z)$.  We use a decoherence-free two-level system as considered in Ref.~\cite{GEMpolariton} to model the storage and retrieval dynamics.  This model can also be applied to a three-level system in the adiabatic limit, which utilizes two ground states coupled by a detuned control field for optical storage \cite{OL}.  With atom-light coupling constant $g$, effective linear atomic density $N$, $\alpha(t,z)$ and $E(t,z)$ are found by numerically solving 
\begin{eqnarray}
\frac{\partial}{\partial t}\alpha(t,z) &=&  -i  \ \eta(t,z)  \ \alpha(t,z)+i g \ E(t,z)  \nonumber \\
\frac{\partial}{\partial z} E(t,z) &=& \phantom{-} i g N \ \alpha(t,z). \label{2lev}
\end{eqnarray}

\begin{figure}
   \includegraphics[width=\columnwidth]{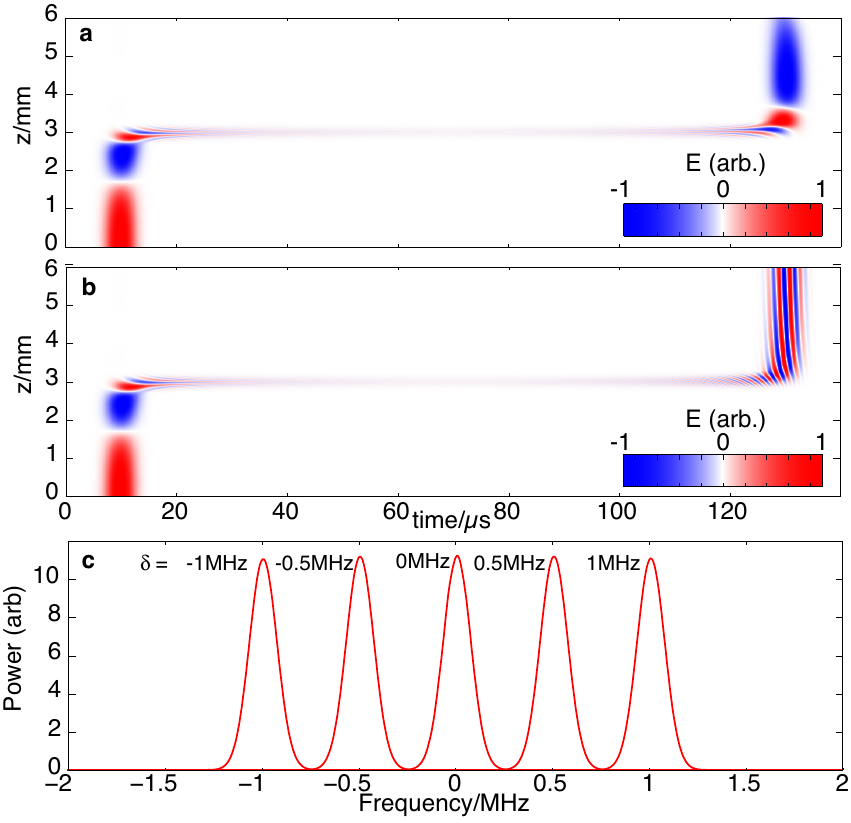}
    \caption{The real part of the optical field when a: $\delta=0$~MHz. b: $\delta=0.5$~MHz.  c: Fourier spectra of the output with $\delta$=-1,-0.5, 0, 0.5 and 1~MHz for Gaussian inputs pulses .  The effective optical depth in these simulations is $gN/\eta=3.75$.}
\label{fshift}
  \end{figure}

A simple frequency shift of the retrieved light can be achieved by adding or subtracting energy to light while it is in storage.  An offset ($\delta$) can be added to the atomic frequency sprectum as it is inverted so that $\eta(z,t)\rightarrow -\eta(z,t) +\delta$.  In this way every frequency component in the pulse is shifted by $\delta$ on retrieval. An example is shown in Fig.~\ref{fshift}.  Panels a and b show the real part of the optical field during storage with $\delta=0$ and  $\delta=0.5$~MHz respectively.  The rapid phase rotation of the output in Fig.~\ref{fshift}b indicates a shift of the output frequency relative to the frame of the simulation that is rotating at the input optical frequency.  Fig.~\ref{fshift}c shows the Fourier spectra of the output pulses correspondingly shifted for five values of $\delta$  between -1 and 1~MHz.   Apart from this being an interesting processing capability, being able to frequency shift in this way enables compensation of inherent frequency shifts that can occur in GEM \cite{HetetPRL} under conditions of short storage time. In the case of three-level GEM a shift can be introduced by changing the frequency of the coupling field that controls the connection between the two atomic ground states.  This may be easier in practice since a precise frequency shift could be dialled up by using an acouto-optic modulator to control the coupling beam frequency in a similar manner to EIT experiments that show freqeuncy shifts \cite{CampbellNJP09}.

\begin{figure}
   \includegraphics[width=\columnwidth]{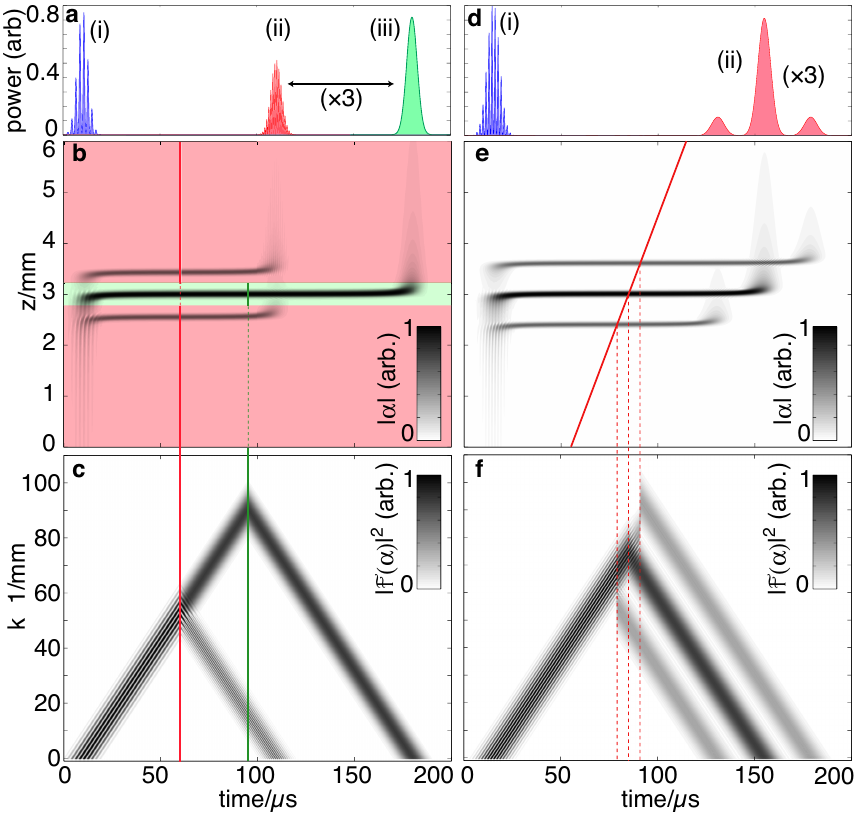}
    \caption{a: Modulated input pulse (i) and recalled pulses showing split sideband (ii) and carrier (iii) light with $\times$3 magnification. b: The atomic polarisation in the $(t,z)$ plane.  In the red (green) shaded regions the atomic frequency gradient is inverted at 60~$\mu$s (70~$\mu$s). c: The spatial Fourier spectrum of $\alpha$. d: Modulated input pulse (i), recalled pulses (ii) showing three separate frequency components with $\times$3 magnification. e: The atomic polarisation in the $(t,z)$ plane. The inversion of $\eta(t,z)$ occurs along the red line given by $60z - 6t - 330=0$. f: The spatial Fourier spectrum of $\alpha$. The effective optical depth in these simulations is $gN/\eta=3.75$.}
\label{fsplit}
  \end{figure}

It is also possible to selectively recall different frequency components. Consider a modulated Gaussian pulse as shown in Fig.~\ref{fsplit}a(i).  When stored in the memory it will give rise to the atomic polarisation shown in  Fig.~\ref{fsplit}(b), which shows the Fourier spectrum of the stored light.  We can identify  a \textit{carrier} at $z=3$~mm and two \textit{sidebands} at $z= 3 \pm 0.4$~mm.   In order to release the different frequency components at different times, we invert $\eta(t,z)$ in segments.  In the left column of Fig.~\ref{fsplit}, $\eta(t,z)$ for $2.8>z>3.2$~mm is inverted at $t=60$~$\mu$s, as indicated by the red line and shading.  In this way the modulation sidebands in the pink shaded area are recalled first, while the carrier is recalled later by inverting the central part of the atomic spectrum at $t=70$~$\mu$s.  The output pulses in Fig.~\ref{fsplit}a show a first peak (ii) which contains the sideband frequencies and a second peak (iii) that contains only the carrier.

Figure~\ref{fsplit}c shows the spatial Fourier spectrum of $\alpha$ given by $|\mathcal{F}(\alpha)|^2$.  As described in Ref.~\cite{GEMpolariton}, this $k$-space picture is a powerful way of visualising the behaviour of the memory and, in fact, conforms to a normal mode description of the coupled atom-light system. The normal mode starts at $k=0$ when the light enters the storage medium. Because the atomic polarisation along the $z$ axis is given by the Fourier transform of the input pulse, the cross-section of the normal mode in the $k$ axis has the shape of the input pulse temporal profile.  The normal mode evolves to higher $k$ values at a rate determined by the gradient of $\eta(t,z)$.  In the case where $\eta(t,z)$ is uniformly inverted, as considered in Ref.~\cite{GEMpolariton}, the normal mode evolves backwards to $k=0$ where the pulse is re-emitted.  In Fig.~\ref{fsplit}c the normal mode is split because the atomic frequency gradient is inverted in two parts. The sidebands form a modulated mode that is emitted at $t=95\mu$s while the carrier forms a second mode that is emitted at $t=180\mu$s

In general, engineering a particular form of $\eta(t,z)$ allows the realisation of arbitrary dispersion profiles.  For example, by switching the sign of $\eta(t,z)$ along the the red line in Fig.~\ref{fsplit}e, we achieve linear dispersion that directly recovers the Fourier spectrum of the input pulse into the time domain. This is seen in the output pulses of Fig.~\ref{fsplit}d(ii). The normal mode picture (Fig.~\ref{fsplit}f) also demonstrates the transformation of the modulated pulse into a frequency spectrum.

 \begin{figure}
   \includegraphics[width=\columnwidth]{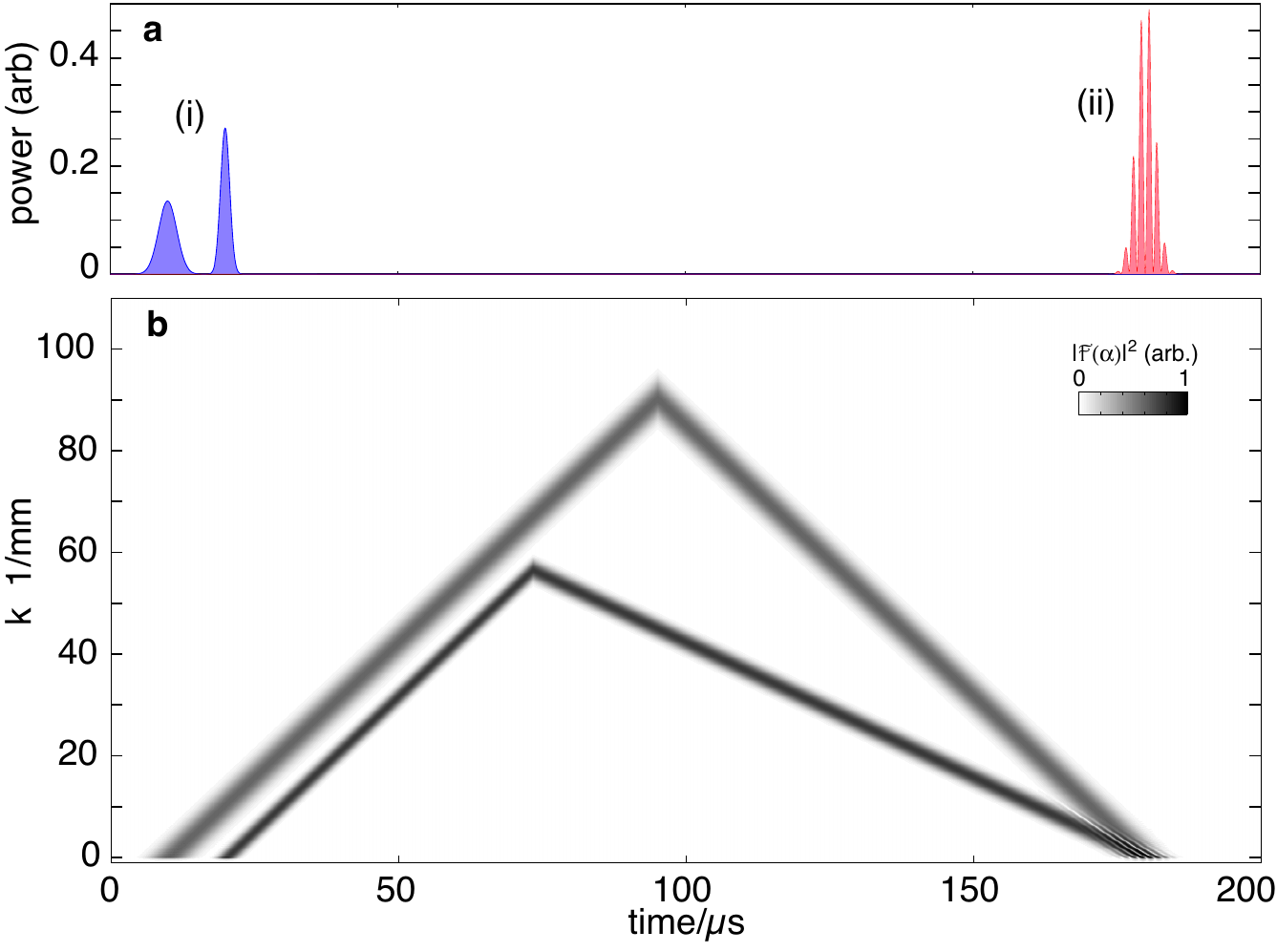}
    \caption{a: (i) Pulses of equal energy but with widths 3.2 and 1.6~$\mu$s and a frequency difference of 3/$\pi$~MHz enter the memory at separate times. (ii) They are recalled with identical pulse width to give 100\% interference visibility.  b: The normal modes of the two pulses. The effective optical depth in these simulations is $gN/\eta=3.75$.}
\label{compress}
  \end{figure}

In our final example we consider a combination of frequency selective recall and spectral compression. Photon echoes can pulse compress via chirp excitation \cite{wangOL2000}.  Although our scheme works in a different manner, pulse expansion (compression) can also be implemented in GEM if the recall gradient used is lower (higher) than that used for writing to the memory \cite{nature09}.  The spread of optical frequencies in the retrieved field will have been reduced (increased), leading to longer (shorter) pulse duration.  Consider the example shown in Fig.~\ref{compress}a(i) of two equal energy pulses, one twice the width of the other and with a relative frequency difference ensuring that the pulses are stored in different parts of the memory, as demonstrated in Fig.~\ref{fsplit}.  We can use GEM to interfere these disparate pulses with 100\% visibility by controlling switching time and retrieval gradient as shown by the recalled pulse of Fig.~\ref{compress}a(ii).  The normal mode picture (Fig.~\ref{compress}b) explains visually how this is done. The shorter pulse is recalled with a gradient reduced by a factor of 2 and switched at $t$=73~$\mu$s.   This leads to a slower recall as seen by the reduced slope of the normal mode in $k$-space. The longer pulse is recalled with an unchanged atomic gradient that is inverted at $t$=95~$\mu$s so that both pulses are recalled from the medium at the same time.

The schemes described in this letter will ultimately be limited by the level structure of the atomic systems used for the memory.  In the case of two-level rare-earth doped solid state systems \cite{HetetPRL}, a Stark shift is used to implement the atomic frequency gradient and this can remain linear over hundreds of GHz.   For ground-state alkali metal memories \cite{nature09}, the bandwidth will be a few GHz limited by the regime of linear ground-state splitting.  Both platforms show potential for long-lived storage.  Coherence times for optical transitions in solid state can be tens of milliseconds \cite{AlexLUM}.  For warm alkali gas systems coherence times are generally limited to a few milliseconds by atomic diffusion \cite{NovikovaPRL07}, although ultra-cold alkali systems can have coherence times reaching hundreds of milliseconds \cite{SchnorrbergerPRL09,ZhangPRL09}. Regardless of the atomic system, one requires fine control over the applied atomic frequency shift. This could be achieved with multi-element electrodes or magnetic coils to introduce tuneable Stark or Zeeman shifts, respectively. In both cases, optimal control will be achieved by using an elongated storage geometry that does not require extremely rapid spatial variations in the electric or magnetic field.  In the case of warm atomic vapour, this will also combat the impact of atomic motion along the length of the storage medium.

We thank J.J. Longdell, M. Hedges and M. Sellars for enlightening discussions. This work was supported by the Australian Research Council.

\normalsize

\end{document}